\documentclass[12pt]{article}%
\nonstopmode

\pdfoutput=1

\usepackage{bbm}

\usepackage{amsmath}
\usepackage{amsbsy}
\usepackage{amscd}
\usepackage{amsopn}
\usepackage{amstext}
\usepackage{amsxtra}
\usepackage{amssymb}
\usepackage{amsthm,amsfonts}
\usepackage{graphicx}
\usepackage{xcolor}
\usepackage{subcaption}
\usepackage{fancybox}
\usepackage{bbm}

\usepackage[square, sort, numbers]{natbib}

\usepackage{centernot}
\usepackage{amssymb}

\usepackage{epsfig}

\newtheorem{theorem}{Theorem}[section]
\newtheorem{corollary}{Corollary}[theorem]

\newtheorem{proposition}[theorem]{Proposition}

\def\beq{\begin{eqnarray}}
\def\eeq{\end{eqnarray}}
\def\beqq{\begin{eqnarray*}}
\def\eeqq{\end{eqnarray*}}
\def\beeq{\begin{eqnarray*}}
\def\eeeq{\end{eqnarray*}}
\def\be{\begin{equation}}
\def\ee{\end{equation}}

\def\BF={{\mathbb{F}}}
\def\BG={{\mathbb{G}}}
\def\BH={{\mathbb{H}}}
\def\bOne={{\bf 1}}

\def\CF{{\cal F}}

\def\var{{\mbox{Var}}}

\def\qed{\hfill$\sqcap\kern-8.0pt\hbox{$\sqcup$}$\\}

\def\BBE{\mathbb{E}}
\def\BBP{\mathbb{P}}

\def\BBR{\mathbb{R}}

\begin{document}
\title{Optimal Portfolios of Illiquid Assets}
\author{T. R.~Hurd$^1$, Quentin H. Shao$^1$, Tuan  Tran$^1$\\
\emph{ $^1$Department of Mathematics \& Statistics,}\\
\emph{McMaster University, Canada }}

\maketitle

\abstract{This paper investigates the investment behaviour of a large unregulated financial institution (FI) with CARA risk preferences. It shows how the FI optimizes its trading to account for {\it market illiquidity} using an extension of the Almgren-Chriss market impact model of multiple risky assets.  This expected utility optimization problem over the set of adapted strategies  turns  out to have the same solutions as a mean-variance optimization over deterministic trading strategies. That means the optimal adapted trading strategy is both deterministic and time-consistent. It is also found to have an explicit closed form that clearly displays interesting properties. For example, the classic constant Merton portfolio strategy, a particular solution of the frictionless limit of the problem, behaves like an attractor in the space of more general solutions.  The main effect of temporary market impact is to slow down the speed of convergence to this constant Merton portfolio. The effect of permanent market impact is to incentivize the FI to buy additional risky assets near the end of the period. This property, that we name the {\it Ponzi property}, is related to the creation and bursting of bubbles in the market. 
The proposed model can be used as a stylized dynamic model of a typical FI in the study of the {\it  asset fire sale channel} relevant to understanding systemic risk and financial stability.  

}
\section{Introduction}
Long after such landmark contributions as the Markowitz mean-variance strategy (\citet{Mark52})  and the Merton portfolio model introduced in \citet{Merton69}, our understanding of optimal portfolio selection has continued to develop. We now have learned how to analyze investment in imperfect markets that have frictions such as transaction costs (\citet{davis1990}, \citet{perold1988implementation}) and  price impact (\citet{AC}, \citet{almgren2003optimal},  \citet{Schoneborn}), and have complex dynamics such as jumps (\citet{cartea2015optimal},   \citet{moazeni2013optimal} and   \citet{Pham}). Indeed, this problem has generated hundreds of research papers. Our goal now is to present a solvable model of optimal investment for a large financial institution (FI) in a many-asset setting. It is based on the expected utility maximization criterion, and it accounts for {\it market illiquidity}, which means the transaction costs to pay and the fact that trades have a permanent price impact. The underlying investment assets, which may be very illiquid,  are assumed to follow  Bachelier dynamics, meaning they are modelled by correlated arithmetic Brownian motions.  For these assumptions to make financial sense, the optimal strategy should be implemented only over a time horizon $[0,T]$ short enough that the Bachelier dynamics remains a reasonable approximation (we take as a benchmark $T=1/2$ years in our examples).  

The class of optimal strategies we obtain has several remarkable properties. First, the general multidimensional problem has a closed-form solution expressible in terms of a matrix-valued equation that can  be efficiently computed with a controllable error.  Second, the solution depends on the full range of important parameters: temporary price impact, permanent impact, risk aversion, the initial portfolio weights, the risk free interest rate, and the parameters underlying the Bachelier dynamics. Thirdly, the optimal strategies, which are a priori adapted processes that solve a version of Merton's problem, turn out to be deterministic over a finite time horizon and to solve a version of the Markowitz mean-variance optimization.   This  property implies that our investment strategies are fully consistent with dynamic programming, despite being deterministic solutions of a time-inconsistent mean-variance optimization problem.

The aim of this paper is to study the effect  of market illiquidity on the behaviour of an FI. {\it Funding illiquidity} (see for example \citet{BrunPede09}) is the distinct effect that the balance sheet of an FI may experience funding shocks caused by unanticipated withdrawals by depositors. To keep the focus of the paper squarely on market liquidity, funding illiquidity is ruled out by the assumption that deposits are constant and sufficient to support all asset purchases of the FI. 

The proposed model and its solution is closely related to some important contributions to the existing literature. Our solutions reduce to the Markowitz optimal portfolios, or equivalently to Merton's optimal solutions,  when permanent and temporary impact are both assumed to be zero. The posed finance problem is inspired by the mean-variance optimal liquidation problem studied by  \citet{AC}, but differs in that there is no constraint placed on the portfolio holdings at the terminal time $T$.   Finally, under certain initial conditions the FI will seek to liquidate a large position, creating what has been called an asset fire sale. Our strategies extend to this setting and give natural criteria similar to those discussed by  \citet{brown2010optimal} that solve the problem of the order in which different assets are liquidated. 


\section{Optimal Portfolio Strategies}

This paper will investigate the investment strategies of a large financial institution (FI) with CARA risk preferences (CARA is short for constant absolute risk aversion that trades continuously over a finite time horizon $[0, T]$ in a market with imperfect liquidity. This is similar to a problem studied in \citet{Zhang}. The changes caused by rebalancing a portfolio of a large FI may amount to a large fraction of the total daily volume traded of these assets and significantly impact these assets' prices. It is well understood that this effect will lead the FI to break large orders into small portions spread over time to reduce market liquidity costs, while still aiming to rebalance its portfolio. By taking  additional time to reduce liquidity costs, the FI now faces additional uncertainty in the price of the assets. To handle this delicate balance between liquidity costs and price uncertainty, the FI will be inclined to consider utility optimization. 

There are sound economic reasons to optimize using an exponential (CARA) utility function:  It leads to a tractable time-consistent strategy where additional information does not provide additional utility, and is similar to the original Mean-Variance optimization of \citet{AC}. Since the strategy is only implemented over $[0, T]$, at time $T$ the FI will update its information and continue in a similar way to rebalance over the subsequent period. This rebalancing is necessary to account for shortcomings of the model, changes in the balance sheet, and  \textit{unanticipated events} that cause fundamental changes to the parameters of the price dynamics. 

A number of simplifications will be assumed about this problem. The total information available to the FI up to any given instant of time $t$ is modelled by a filtration $\{\CF_t\}_{t\ge 0}$ on a given probability space $(\Omega,\CF,\mathbb{P})$. The market consists of  one risk-free asset with zero interest rate, and $d$ risky assets whose {\it true} price process is $S_t=(S^{(1)}_t,...,S^{(d)}_t)'$ and whose {\it transaction} price process  is $\tilde S_t=(\tilde S^{(1)}_t,...,\tilde S^{(d)}_t)'$.  Here and in the following, we adopt matrix notation where $M'$ denotes the matrix transpose of $M$. Let us denote the vector of the amounts held in risky assets by $(q_u)_{u\in[t, T]}$   and the vector of trading rates of the large trader by $v_u:=\dot{q}_u:=dq_u/du, u\in[t, T]$.  

Like \citet{AC} and others, we suppose that the price of risky assets follows a $d-$dimensional Bachelier model with both linear permanent and linear temporary market impact (parametrized by $\Lambda$ and $\Gamma$ respectively):  
\begin{eqnarray}
dS_t&=& (\Lambda v_t+\mu)\ dt+\Sigma\ dB_t\ ,\nonumber\\
\tilde{S}_t&=&S_t+\Gamma v_t\ .
\label{impactmodel}
\end{eqnarray}
Here,  $B_t$ is a $d-$dimensional Brownian motion and $\Sigma\in\mathbb{R}^{d \times d}$ is the volatility matrix. The drift term $\mu=b-d+\Lambda Q\in\mathbb{R}^{d} $ is assumed to be constant. It takes into account the trending rate $b$, dividend rate $d$ and  aggregated permanent market price impact due to external traders $Q$.  

A more general formulation of the model that does not require linear market impact is certainly possible, and will not change many of the same basic properties. However, the assumption of  linear  impact leads to significantly more tractable optimal strategies. Moreover, as shown by \citet{Gatheral}, so-called {\it  dynamic arbitrage} is ruled out by choosing the permanent impact to be linear. It is further assumed that the permanent and temporary impact matrices $\Lambda$ and $\Gamma$ are symmetric and non-negative definite. The assumption that $\Gamma$ is symmetric is without loss of generality.  On the other hand, $\Lambda$ is assumed to be symmetric not for economic reasons but for convenience: when it has an anti-symmetric part, a somewhat more complicated explicit solution is obtainable.  Models similar to ours have been studied by \citet{almgren2007adaptive}, \citet{Gatheral} and \citet{schied2009risk}. We refer to the review by \citet{HurdShaoTuan16a} for further background and justification of these and other similar models.

\subsection{The Merton Problem } Merton's problem, introduced in \citet{Merton69}, aims to determine the strategies followed by utility optimizing investors in continuous time market models.   To this end, we now consider the most general portfolio strategy, or control process, that trades within the market impact model \eqref{impactmodel} over some time interval $[s,t]\subset \mathbb{R}_+$. In our setting, each possible strategy will be simply a $d-$ dimensional trading rate process $v=(v_u)_{u\in[s,t]}$ that is adapted to the information filtration $\{\CF_t\}$: We denote the set of such 
{\it admissible} strategies by $\Pi^{ad}[s,t]$. The subclass of {\it deterministic} strategies where each value $v_u, u\in[s,t]$ is $\mathcal{F}_s$ measurable  is denoted by $\Pi^{det}[s,t]$. 

Given any control process $v\in\Pi^{ad}[0,s]$ for $0<t<s$, the {\it cash net of debt owed} $C_t:=C^v_t$ and marked-to-market {\it equity}, or {\it assets net of debt owed}, $X_t:=X^v_t:=C_t+q_t'S_t$ are given by:
\begin{eqnarray}
\label{C_SDE}C_t&=&C_0-\int_0^t v_u' \tilde{S}_u\ du=C_0-\int_0^t v_u' S_u\ du-\int_0^tv_u'\Gamma v_u \ du\ ,\\
\label{X_SDE}X_t&=&X_0+\int_0^tq_u\ dS_u-\int_0^tv_u'\Gamma v_u\ du\ .
\end{eqnarray}
where the second equation is obtained by integration by parts. Note that here and henceforth, the superscript $v$ that labels processes controlled by $v$ will be omitted. 

The interpretation of \eqref{C_SDE} and \eqref{X_SDE} in terms of the firm's balance sheet is that assets are stochastic due to fluctuations of $S$, while the debt, thought of as deposits, is assumed to be constant and sufficient to fund all trades. In other words, we focus on market illiquidity without funding illiquidity.  It is consistent with the Principle of Limited Liability that a firm becomes insolvent when its equity $X_t$ becomes negative. In the following,  an insolvent firm with negative equity $X_T<0$ at a time $T$, will  be declared to be {\it in default}, implying that the laws of bankruptcy will be applied to the firm. 

The FI can now try to solve Merton's optimal problem of a CARA investor with constant absolute risk aversion parameter $\lambda>0$ over any period $[t,T]$. For each $t$, they may express the value function $J_t$ achieved in terms of a {\it certainty equivalent value} $W_t$,
 \be\label{Problem1}  J_t:=-e^{-\lambda W_t}:=\text{sup}_{v\in\Pi^{ad}[t,T]}\  \left(-\mathbb{E}[-e^{-\lambda X_T}|\CF_t]\right)\ .
\ee
If the supremum exists, it is achieved by adopting an {\it optimal control} denoted by $v^*(t)=(v_u^*(t))_{u\in[t,T]}$, which will be an adapted process over $[t,T]$.
The CARA investment problem in general always satisfies the {\it dynamic programming principle} (see  \citet{schied2010optimal}), which means that for any $s\le t\le T$,  $v_u^*(t)=v_u^*(s)$ for all $u\ge t$ and 
\be\label{DPP1} -e^{-\lambda W_s}=\text{sup}_{v\in\Pi^{ad}[s,t]}\ \left(-\mathbb{E}[e^{-\lambda W_t}|\CF_s]\right)\ .
\ee

An investor restricted to deterministic strategies over $[t,T]$ cannot achieve a higher certainty equivalent value than equation \eqref{Problem1}. Therefore,  if  $\widetilde W_t$ is defined by   
\be\label{Problem2} -e^{-\lambda \widetilde W_t}:=\text{sup}_{v\in\Pi^{det}[t,T]}\  \left(-\mathbb{E}[-e^{-\lambda X_T}|\CF_t]\right)
\ee
then $\widetilde W_t\le W_t$. The first result of this paper, stated next,  is that \eqref{Problem1} is always optimized by deterministic strategies and therefore $W_t=\widetilde W_t$ for  $t\ge 0$. Moreover, it will be found in subsequent sections that the optimal control and value functions can be expressed in closed forms  involving one-dimensional integrals that solve a system of ordinary differential equations of Riccati type. First, however, a note about notation: Because $v^*$ and $q^*$  turn out to be deterministic, we henceforth replace the stochastic process notation $v_u$ by function notation $v(u)$ and moreover suppress the dependence on the investment period $[t,T]$.

\begin{theorem}\label{Optimal} Under the above modelling assumptions, there is a (possibly infinite) maximal time  $T^*\in\BBR_+\cup\{\infty\}$) such that for any finite time horizon $[t,T]$ with  $0\le t\le T\le T^*$:
\begin{enumerate}
  \item 
  The optimal strategy $v^*(u), u\in[t, T]$  exists, is unique and $\CF_t$ measurable, hence deterministic.
  \item The value function $\widetilde W_t$  achieved over $[t,T]$, when restricted to deterministic strategies, equals $W_t$. 
  \item The value function has the form $W_t=X_t+V(T-t, q)$ where $V(\tau, q), \tau=T-t$ solves the non-linear partial differential equation
  \be\label{MV_optimal}
 \hspace{-.75in} -\partial_\tau V+q'\mu-\frac{\lambda}{2} q'\Sigma\Sigma' q+\frac 14 (\Lambda q+\partial_qV)'\Gamma^{-1}(\Lambda q+\partial_qV)=0,\quad V(0,q)=0\ 
  \ee
  on the domain $[0,T]\times \BBR^d$.
\item Given initial holdings $q$ at time $t$, the optimal portfolio holdings $q^*(u)$ for    $u\in [t, T]$ solves the system of ODEs:
\beq\label{optimalq}
\frac{dq}{du}=\frac{\Gamma^{-1}}{2}\bigl(\partial_q V(T-u,q)'+\Lambda q\bigr)\ ,\quad q(t)=q\ .
\eeq

\end{enumerate}
\end{theorem}

\bigskip The proof of this theorem is found in the Appendix. As we shall see in Section \ref{Explicit}, $V(\tau, q)$ is a quadratic form in $q$ with time-dependent coefficients and thus the  ODE \eqref{optimalq} for $q^*$ is linear and can be solved explicitly.

\subsection{Mean, Variance, Probability of Default and Time Consistency}
From equations \eqref{C_SDE} and \eqref{X_SDE} we can deduce that, if $v$ is deterministic, then for any $0\le s\le t\le T$, the equity $X_t$ conditioned on $\CF_s$  is normally distributed with mean and variance given by
\begin{eqnarray}
\mathbb{E}[X_t|\CF_s]&=&X_s+ \int_s^t\Bigl(q'(u)(\Lambda v(u)+\mu)- v'(u)\Gamma v(u)\Bigr)\ du, \\ \var[X_t|\CF_s]&=&\int_s^t\ q'(u)\Sigma\Sigma'q(u)\ du\ .
\end{eqnarray}
In particular, the fact that $X_T|\CF_t$ is always normal implies that \be
\BBE[e^{-\lambda X_T}|\CF_t]=e^{-\lambda\left(\mathbb{E}[X_T|\CF_t]-\frac\lambda 2 \var[X_T|\CF_t]\right)}
\ee and hence from \eqref{Problem2} and Theorem \ref{Optimal} one deduces that
\be\label{MV_optimal2} W_t=\widetilde W_t=\text{sup}_{v\in\Pi^{det}[t,T]} \left(\mathbb{E}[X_T|\CF_t]-\frac\lambda 2 \var[X_T|\CF_t]\right)\ .
\ee
This demonstrates the well-known equality of the certainty equivalent value for CARA optimization with the value function for Markowitz' mean-variance (M-V) optimization, as well as the coincidence of their optimal strategies, when the optimal equity processes under consideration are all normally distributed.

In practice, the firm's {\it default probability} (DP), meaning the probability that $X_T<0$, may be preferable to variance as a risk measure for institutional investors, as it gives more information about  bad scenarios that need to be controlled. In the Bachelier model, the normality that follows for deterministic strategies implies that  over any time horizon $[t,T]$, the Mean-Variance (M-V) criterion\\ 

\bigskip
\noindent\textbf{Problem M-V}
\begin{eqnarray}
\label{MV}
W_{V}(t, T, q, x,E)&:=&\text{min}_{v\in\Pi^{det}[t,T]}\ \text{Var}[X_T|\CF_t]  \\
&&\text{subject to } \BBE[X_T|\CF_t]=E \ ,\nonumber
\end{eqnarray}
and the Mean-Default Probability (M-DP) criterion \\

\bigskip
\noindent\textbf{Problem M-DP}
\begin{eqnarray}
\label{MDP}
W_{DP}(t, T, q, x,E)&:=&\text{min}_{v\in\Pi^{det}[t,T]}\ \BBP[X_T< 0|\CF_t]  \\
&&\text{subject to } \BBE[X_T|\CF_t]=E\ ,\nonumber
\end{eqnarray}
are both solved by the same optimal trading strategy when $\BBE[X_T|\CF_t]=E>0$. This is because $\BBP[X_T< 0|\CF_t]$ is strictly increasing in $\text{Var}[X_T|\CF_t]  $ as long as  $\BBE[X_T|\CF_t]>0$ is fixed.  Moreover, if $v^*(\lambda)$ denotes the optimizer of \eqref{MV_optimal2}, and $X^*_T(\lambda)$ is the optimal equity it achieves, then $v^*(\lambda)$ also optimizes problems \eqref{MV} provided  $E=E(\lambda):=\BBE[X^*_T(\lambda)|\CF_t]$, and also  \eqref{MDP} if in addition $E(\lambda)>0$ .

Since Merton's optimal problem \eqref{Problem1} satisfies Bellman's Dynamic Programming Principle at all times, its optimal strategies are  ``time-consistent'',  which means that the optimal strategies computed for any two periods $[t,T]$ and $[s,T]$  always coincide on the intersection $[s\vee t, T]$. On the other hand, it is known (\citet{}) that mean-variance optimization is generally time-inconsistent and optimal {\it adapted} strategies starting at one time do not usually appear optimal at a later time. Surprisingly, Theorem \ref{Optimal} 
combined with equation \ref{MV_optimal2} implies that in the present context, both the mean-variance and mean-default probability problems \eqref{MV} and \eqref{MDP}  are in fact time consistent, provided the optimization is restricted to deterministic strategies. The following result summarizes these relationships. 

\begin{corollary}\label{MVMDPequivalence}
For any fixed time horizon $[t,T]$, let $E(\lambda)=\BBE[X^*_T(\lambda)|\CF_t]$ be the expected value of equity computed for the optimal strategy $v^*(\lambda)$ of the CARA investment problem \eqref{MV_optimal2} with risk aversion parameter $\lambda$. Let $\underline{E}$ and $\overline{E}$  be the infimum and supremum  of  $E(\lambda)$  when $\lambda$ varies over $[0, \infty)$.  Then:
\begin{enumerate}
  \item For any $E\in (\underline{E}, \overline{E}),$ there exists a unique $\lambda=\lambda(E)$ such that $E(\lambda)=E$. 
   \item For all possible values of  $E(\lambda)$, the deterministic optimal strategies $v^*$ of Problem M-V coincide with the unique adapted optimal strategy $v^*(\lambda)$ of \eqref{MV_optimal2}. 
  \item The optimal strategies computed for any two periods $[t,T]$ and $[s,T]$ are  time-consistent, meaning they coincide on the intersection $[s\vee t, T]$.  
\item If $E(\lambda)>0$, the deterministic optimal strategies $v^*$ of Problem M-V and Problem M-DP also  coincide with each other.  

\end{enumerate}\end{corollary}
\section{Explicit Optimal Strategies}\label{Explicit}
We now exploit the tractability of Merton's problem in the market impact setting to obtain closed formulas (involving matrix algebra and one dimensional integration)  for the optimal trading curve of the financial institution (FI). The techniques invoked in this section are closely related to the methods developed for the optimal liquidation problem in  \citet{}.

\begin{proposition}\label{RicattiEqn} Under the modelling assumptions of Theorem \ref{Optimal}, for any finite $T$ with $T\le T^*$,
 the value function $W_t:=W(t, T, q, x)=x+V(\tau,q), \tau=T-t$  for any $t\in[0,T]$ has the form 
  \be\label{Vform} V(\tau, q)=q'A(\tau)q +B(\tau)q +C(\tau)
  \ee
  where $A,B,C$ are matrix valued functions of dimension $[d,d]$, $[1,d]$ and $[1,1]$ respectively with $A$ symmetric. These functions satisfy Riccati-type ODEs for $\tau>0$:  \beq
 \frac{\partial A}{\partial \tau} &-&(A+ \Lambda/2)'\Gamma^{-1}(A+\Lambda/2)+\frac{\lambda}{2}\Sigma\Sigma'=0,\quad A(0)=0  ,\label{riccatiA} \\
 \frac{\partial B}{\partial \tau} &-& B\Gamma^{-1}(A+\Lambda/2)-\mu'=0,\quad B(0)=0  ,\label{riccatiB}\\
 \frac{\partial C}{\partial \tau} &-&  \frac 14B\Gamma^{-1}B'=0,\quad C(0)=0 . \label{riccatiC}
 \eeq
    \end{proposition}

The next theorem, whose proof is given in Appendix A, solves this system of Riccati equations in closed form  in terms of $E:=\Gamma^{-1/2}(\Lambda/2)\Gamma^{-1/2}$ and the symmetric square root $D$ of $$D^2:=\frac{\lambda}{2}\Gamma^{-1/2}\Sigma\Sigma'\Gamma^{-1/2}\ .$$ It also provides a closed form for the optimal strategy $q^*$. 
\begin{theorem}\label{closedform}
\begin{enumerate}
   \item
   The  solution of the system of Riccati equations \eqref{riccatiA}--\eqref{riccatiC} over the maximal interval $[0,T^*]$ is given by\beq
&&A(\tau)=\Gamma^{\frac{1}{2}}\Bigl(V(\tau)U(\tau)^{-1}-E\Bigr)\Gamma^{\frac{1}{2}}\\
&&B(\tau)=\overline{\mu}'\bigl(E-V(\tau)\bigr)U^{-1}(\tau)\Gamma^{\frac{1}{2}}\\
&&C(\tau)=\frac 14\overline{\mu}'\left(\int_0^\tau\bigl(E-V(s)\bigr)\bigl(U'(s)U(s)\bigr)^{-1}(E-V'(s))ds\right)\overline{\mu}
\eeq
where $\overline{\mu}:=D^{-2}\Gamma^{-1/2}\mu$ and the
 matrix valued functions $U,V$ are given by
\beq
U(\tau)&&=\ {\cosh}(D\tau)-D^{-1}\ {\sinh}(D\tau)E\\
V(\tau)&&=-\ {\sinh}(D\tau)D+\ {\cosh}(D\tau)E.
\eeq
\item 
The maximal time horizon $T^*$ is 
\[ T^*=\ {\inf}\{\tau>0:U(\tau) \mbox{ is not invertible}\}\ .
\]
$T^*$ is finite if $D<E$ and $\infty$ if $D> E$.
\item 
For any $(t, T, q, x),$ the optimal trading curve $q^*(u)$ over the period $[t,T]$ is  
\beq\label{qformula}
q^*(u)&=&\Gamma^{-1/2}U(T-u) U^{-1}(T-t)\Gamma^{1/2}q\\
&&+ \frac12\Gamma^{-1/2}U(T-u)  \int^u_t U^{-1}(T-r)\Gamma^{-1/2} B'(T-r)dr\eeq
 \item For any $(t, T, q, x),$ the  expected value and variance of the optimal terminal equity are:
 \beq 
\label{Opt_E} \BBE[X_T^*(\lambda)|\CF_t]&=&x+q'\Bigl(A(T-t)-\frac{\lambda}{2} L(T-t)\Bigr)q\\\
&&\hspace{-.7in} + \ \Bigl(B(T-t)-\frac{\lambda}{2} M(T-t)\Bigr)q+C(T-t)-\frac{\lambda}{2} N(T-t)\ ,\nonumber\\
 \label{Opt_V} \var[X_T^*(\lambda)|\CF_t]&=&q'L(T-t)q+M(T-t)q+N(T-t)\ ,
 \eeq
\end{enumerate}
where formulas for $L, M, N$ are given in Appendix A.
\end{theorem}

In the special case when $D$ and $E$ are commuting matrices, these formulas decouple into $d$ one-dimensional problems, each of which is similar to the single risky asset case we next discuss.

\subsection{The Case of a Single Risky Asset} 
In the single risky asset case, one can verify that the scalar functions $A,B,C$ and the optimal trading strategy $q^*(u)$ have comparatively simple formulas obtained by reducing those given in Theorem \ref{closedform}. Notice that several distinct possibilities are determined by the relation between $D=\Sigma\sqrt{\frac{\lambda}{2\Gamma}}$ and $E=\frac{\Lambda}{2\Gamma}.$  
\begin{proposition}
\label{singleassetprop}
In the single asset case, 
\begin{enumerate}
  \item When $D>E$ or $2\lambda\Gamma\Sigma^2>\Lambda^2$, Denote $K=\tanh^{-1}(E/D)$  we have $U(\tau)=\frac{\cosh(D\tau-K)}{\cosh K}$ The formulas can be rewritten  in terms of hyperbolic functions as follows 
  \beq
  A(\tau)&=&-\Gamma D\left[\tanh(D\tau-K)+\tanh K\right]\\
 B(\tau)&=&\frac{\mu}{D}\left(\frac{\sinh K}{\cosh(D\tau-K)}+\tanh(D\tau-K)\right)\\
 C(\tau)&=&\frac{\mu^2}{4\Gamma D^3}\left[(\sinh^2(D\tau-K)-1)(\tanh(D\tau-K)+\tanh K)-D\tau\right]\\
 &+&\frac{\mu^2}{2\Gamma D^2}(\tanh K-\frac{\sinh K}{\cosh(D\tau-K)}).
  \eeq
  The optimal trading strategy is given by
\beq  
  q(u)^*&&=\frac{\cosh(D\tau_u-K)}{\cosh(D\tau_t-K)}q+\frac{\mu}{2\Gamma D^2}\left(1-\frac{\cosh(D\tau_u-K)}{\cosh(D\tau_t-K)}\right)\\
  &&+\frac{\mu\sinh K\cosh(D\tau_u-K)}{2\Gamma D^2}\left(\tanh(D\tau_u-K)-\tanh(D\tau_t-K)\right),
  \eeq
  where $\tau_s=T-s.$
   \item When $D=E$ or $2\lambda\Gamma\Sigma^2=\Lambda^2$
   \beq
  A(\tau)&=&0\\
 B(\tau)&=&\frac{\mu}{D}\left(e^{D\tau}-1\right)\\
 C(\tau)&=&\frac{\mu^2}{2D\lambda\Sigma^2}\left[\frac{1}{2}e^{2D\tau}-2e^{D\tau}+D\tau+\frac{3}{2}\right].
  \eeq
  \item When $D<E$ or $0<2\lambda\Gamma\Sigma^2<\Lambda^2$, Denote $K=\coth^{-1}(E/D)$  we have $U(\tau)=-\frac{\sinh(D\tau-K)}{\sinh K}.$ The formulas can be rewritten  in terms of hyperbolic functions as follows 
  \beq
  A(\tau)&=&-\Gamma D\left[\coth(D\tau-K)+\coth K\right]\\
 B(\tau)&=&\frac{\mu}{D}\left(\frac{-\cosh K}{\sinh(D\tau-K)}+\coth(D\tau-K)\right)\\
 C(\tau)&=&-\frac{\mu^2}{4\Gamma D^3}\left[(\cosh^2(D\tau-K)+1)(\coth(D\tau-K)+\coth K)+D\tau\right]\\
 &+&\frac{\mu^2}{2\Gamma D^2}(\coth K+\frac{\cosh K}{\sinh(D\tau-K)}).
  \eeq
  The optimal trading strategy is given by
\beq  
  q(u)^*&&=\frac{\sinh(D\tau_u-K)}{\sinh(D\tau_t-K)}q+\frac{\mu}{2\Gamma D^2}\left(1-\frac{\sinh(D\tau_u-K)}{\sinh(D\tau_t-K)}\right)\\
  &&-\frac{\mu\cosh K\sinh(D\tau_u-K)}{2\Gamma D^2}\left(\coth(D\tau_u-K)-\coth(D\tau_t-K)\right),
  \eeq
  where $\tau_s=T-s.$

   \item When $\lambda=0$, we have $U(\tau)=1-E\tau$ and $V(\tau)=E.$ Moreover
   \beq
 A(\tau)&&=\frac{\Lambda}{2}(\frac{1-U(\tau)}{U(\tau)})\\
B(\tau)&&=\frac{\mu\Gamma}{\Lambda}(\frac{1-U(\tau)^2}{U(\tau)})\\
C(\tau)&&=\frac{\mu^2\Gamma^2}{6\Lambda^3}\left(\frac{-U(\tau)^4+6U(\tau)^2-8U(\tau)+3}{U(\tau)}\right).
\eeq
The optimal trading strategy is given by
\beq
q(u)^*=U(\tau_u)\left(\frac{q}{U(\tau_t)}+\frac{\mu}{4\Gamma E^2}\left(U(\tau_t)+\frac{1}{U(\tau_t)}-U(\tau_u)-\frac{1}{U(\tau_u)}\right)\right).
\eeq
   \end{enumerate}
   \end{proposition}

The third and fourth cases are the cases where $T^*<\infty$, and one finds the solutions become unbounded:  $\lim_{\tau\rightarrow T^*}A(\tau)=\infty$. In cases 1 and 2, the solutions are bounded for all $\tau$, and $T^*=\infty$. 

\subsection{Small Perturbations from Merton's Solution}\label{perturb}
In his original paper \citet{Merton69}, Robert Merton presented the exact solution to the problem of  optimal investment in a frictionless market for an asset price that follows a geometric Brownian motion.  His solution technique also leads to an exact solution of our present model in the limit of zero market impact,  $\Lambda=0, \Gamma=0$, which we will  call the ``Merton solution''. It is of some interest to consider the explicit general solution from the previous section  as a perturbation of the Merton solution, and to investigate the nature of its convergence as market impact goes to zero. We suppose that  $\Lambda=\epsilon\Lambda_1, \Gamma=\epsilon\Gamma_1$ with small $\epsilon$ and denote by $W(t, T,x, q, \epsilon)$ the  certainty equivalent value function  with its dependence on $\epsilon$. For simplicity, we confine our attention to the single asset case of the previous section.


The Merton solution over the period $[t,T]$ with initial conditions $X_t=x, q_t=q$ involves an instantaneous trade that incurs no trading cost, to the optimal value $q^M:=\frac{\mu}{\lambda\Sigma^2}$. This portfolio is then held constant. One can show that this strategy achieves the certainty equivalent value function  $W(t, T, x, q, 0)=x+\frac{\mu^2(T- t)}{2\lambda\Sigma^2}$ which we note is independent of $q$. 

Now, for small $\epsilon$, the general solution of our model is given by Case 1 of Proposition \ref{singleassetprop}, which leads to the following perturbative expansion
 \beq
 W(t, T,x,q,\epsilon)=W(t,T,x,q,0)+\epsilon^{1/2}Q(q)+o(\epsilon^{3/2}).
 \eeq
 where  
  \beq
  Q(q)&:=&\Gamma_1D_1q^2+\frac{\mu}{D_1}q+(2E-1)D_1.
  \eeq
  Here we define $D_1=\Sigma\sqrt{\frac{\lambda}{2\Gamma_1}}$ which does not depend on $\epsilon$ and we have $D=\epsilon^{-1/2}D_1\to\infty$ as $\epsilon\to 0.$ It is obvious that $E$ does not depend on $\epsilon$ either.
   Thus the value function of our problem converges to the value function of the Merton solution with rate of convergence $\epsilon^{1/2}.$
  
   The optimal holding at the terminal time is given by
 \beq
 q_T^*=q^M+(q_t-q^M)\frac{\cosh K}{\cosh(D\tau-K)}+Eq^M\frac{1-\tanh(D\tau-K)}{D(1-\tanh^2K)}.
 \eeq
 
Here $\tau:=T-t.$ It is straightforward that $\lim_{\epsilon\to 0}K=0,$ hence $\lim_{\epsilon\to 0}q_T^*=q^M.$

Let $\tilde{A}(\tau):=\frac{U(\tau)}{V(\tau)}=D\ {\tanh}(D\tau-K),$ the trading rate at the initial time $t$ is given by 
 \begin{eqnarray}
 v_t&&= \lim_{s\to t}\dot{q}_s\\
 &&=q_t\tilde{A}(\tau)+U(\tau)\frac{\mu}{2\Gamma D^2}[\frac{E(D^2-(\tilde{A}(\tau))^2)}{D^2-E^2}-\frac{\tilde{A}(\tau)}{U(\tau)}]\\
 &&=(q_t-q^M)D\tanh(D\tau-K)+\frac{\mu E \cosh K}{2\Gamma_1D_1^2\cosh(D\tau -K)}.
 \end{eqnarray}

 Note that $\lim_{\epsilon\to 0}{\tanh}(D\tau-K)=1$ and $\lim_{\epsilon\to 0}{\cosh}(D\tau-K)=\infty$, we have $\lim_{\epsilon\to 0}v_t=\pm\infty$ depending on if $q_t>q^M$ or $q_t<q^M,$ i.e. the optimal strategy is to trade rapidly in the beginning. We then conclude that the optimal trajectory $q^*(u,\epsilon)$ converges to an $L$--shaped or $\Gamma$--shaped curve when the market impact tends to zero.

This result implies that when market impact is low, the firm will follow an optimal trading strategy very close to the constant holding strategy of the Merton problem. A more surprising fact is the portfolio which starts at the Merton portfolio will remain constant if permanent impact has $\Lambda_1=0$, and all strategies regardless of initial portfolios will move towards the Merton portfolio for sometime initially. In the following subsection, we will show similar results for the Multi-Asset case.

 \section{Numerical Investigations}\label{NumInv}
 
 We now consider the investment behaviour of a hypothetical unregulated financial institution, such as a hedge fund or mutual fund. The firm trades a single risky asset, with initial price $S_0=\$100$, in a market with a $0\%$ risk free rate of return. They use our CARA optimal investment model to trade over non-overlapping half-year trading periods: we focus here on the period $[0,T], \ T=1/2$. The  CARA risk aversion parameter $\lambda$ is chosen to be consistent with a target default probability of $1\%$ for each period. Thus the firm will trade aggressively to maximize their expected return with a quite high tolerance to the potential of default.

 The calibrated parameters of the model given in Table \ref{tab:Parameters} are taken to be fixed at the beginning of the period $t=0$. Note that the firm uses the Bachelier model only for a short period, and expects to recalibrate at the beginning of the each successive period. Since the risky asset is illiquid, there is market impact related to the velocity of trading and the total amount traded: these are assumed to give the temporary and permanent market impact parameter estimates  $\Gamma=\$10^{-7} \text{years/ (units traded)}^2$ and $\Lambda = \$4*10^{-8}/\text{unit}$.

Balance sheets for a small, medium and large firm will be considered, all with a risky asset-to-equity ratio of $4:1$. The initial stock holdings are $q_0=[50000,200000,800000]$ from which the initial firm equity and cash net of debt are then determined to be $X_0=0.25*q_0*S_0, C_0=-0.75*q_0*S_0$ . In all three cases, as indicated above,  the financial institution targets a fixed  default probability under the optimal strategy. This is implemented by choosing the internal risk aversion parameter $\lambda$ so that the $W_{DP}(0, T, q_0, X_0,E(\lambda))=0.01$. Note that even though the optimal strategy does not depend on $X_0$ for fixed $\lambda$, this specification of $\lambda$ depends on $X_0$. Thus firms that differ only in $X_0$ do adopt different investment strategies.

 \begin {table}[h]
\caption {Benchmark Parameters} \begin{center}
\begin{tabular}{rl}
\bf Calibrated Parameter   & \bf   Model Parameter Value\\
\hline
Initial Stock Price & $S_0=\$100$ \\
Trading Period & $[0,T], \ T=0.5$ year\\
$20\%$ Annualized Volatility &$\Sigma=\$20/\text{unit}/\sqrt{\text{year}}$\\
$4\%$ Annual Growth & $\mu=$4/unit/year\\
Temporary Market Impact & $\Gamma = \$10^{-7} \text{year/ (unit)}^2$\\
Permanent Market Impact & $\Lambda = \$4*10^{-8}/\text{unit}$\\
 $\lambda$ such that Probability of Default $=0.0005$  & $\lambda$ varies\\
Initial Holdings & $q_0=[50000,200000,800000]$\\
Initial Cash net Debt owing & $C_0=-0.75*
q_0*S_0$\\  
Initial Equity& $X_0=0.25*q_0*S_0$
\label{tab:Parameters}
\end{tabular} 
\end{center}
\end{table}
 
 \subsection{The Efficient Frontier}
 
\begin{figure}[h]
\includegraphics[width=0.8\columnwidth]{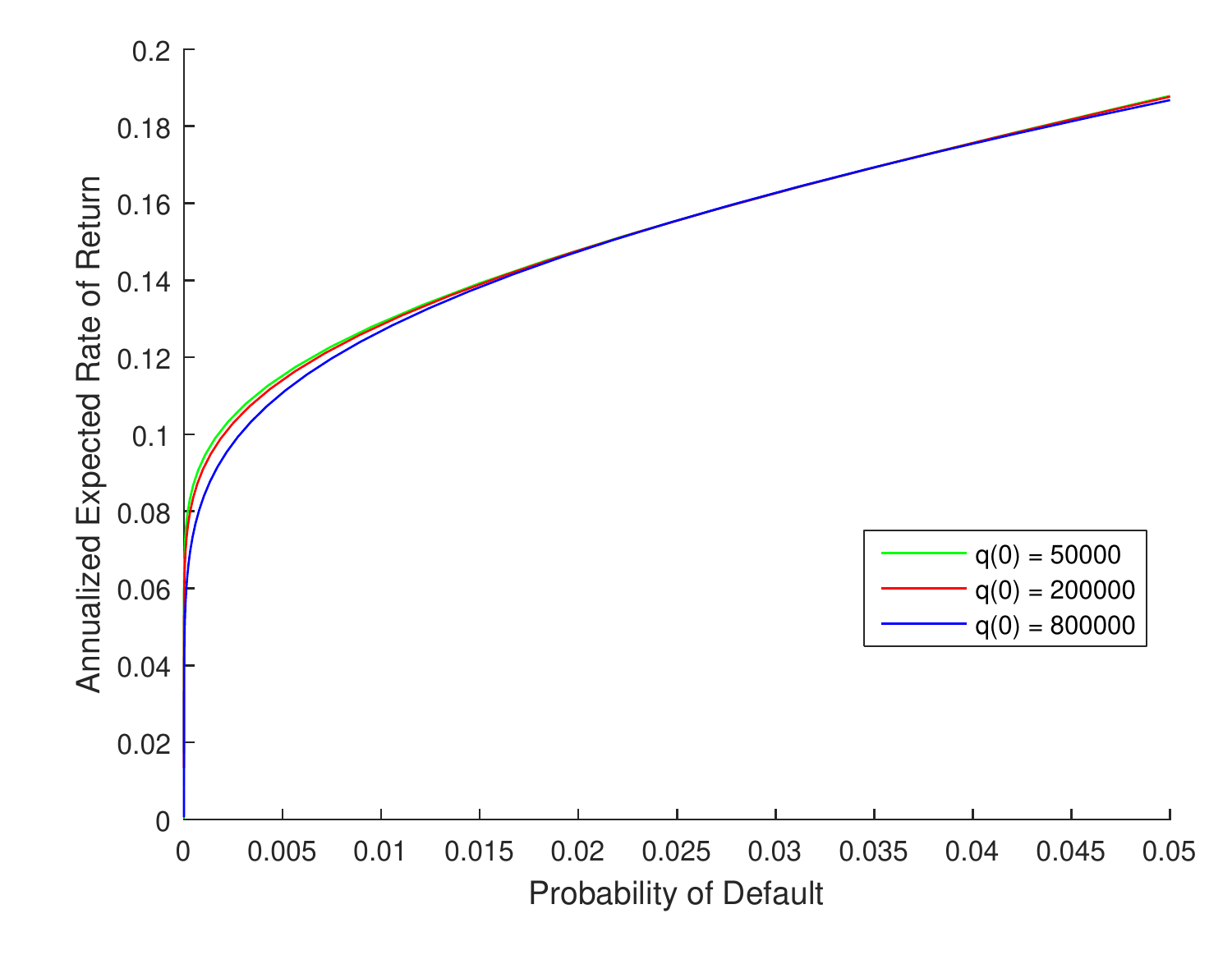}
\caption{ The efficient frontier for three firms with parameters given in Table \ref{tab:Parameters}, showing their default probability and expected rate of return on equity, when adopting their optimal portfolio with risk aversion parameters $\lambda$ varying over $[0,\infty)$.}
\label{EF_plot}
\end{figure}

Figure \ref{EF_plot} shows for each of the three firms how the expected rate of return on equity (ERR) and default probability (DP) for their CARA/MV/DP optimal strategies depend as $\lambda$ varies over the set of feasible values $[0,\infty)$. These quantities are computed by the formulas 
\be {\rm ERR}(\lambda)=\frac{1}{T}(\frac{E(\lambda)}{X_0}-1)\ ,\quad
{\rm DP}(\lambda)=N\left(-\frac{E(\lambda)}{\sqrt{V(\lambda)}}\right)
\ee
where $N(\cdot)$ denotes the CDF of the standard normal  and $E(\lambda), V(\lambda)$ are given by \eqref{Opt_E} and \eqref{Opt_V} . Such a graph is called an {\it efficient frontier}, and it summarizes the results a firm may achieve by adopting  different possible risk aversion parameters.

As explained earlier, the three firms each select the optimal investment strategy given by the value $\lambda$ that leads to ${\rm DP}(\lambda)=0.01$: with the benchmark parameters given in Table \ref{tab:Parameters}, the three values they compute are $\lambda=[ 2.56 \times 10^{-7}  , 6.7 \times 10^{-8} , 1.83 \times 10^{-8} ]$. While Figure \ref{EF_plot} suggests that, ceteris paribus, larger firms have a lower efficient frontier, this ordering can be made to reverse by increasing the permanent impact parameter.


 \subsection{Properties of the Optimal Trading Curve}
 \label{StratProps}
\begin{figure}[h]
\centering
\begin{subfigure}[h]{0.45\textwidth}
\includegraphics[width=1\columnwidth]{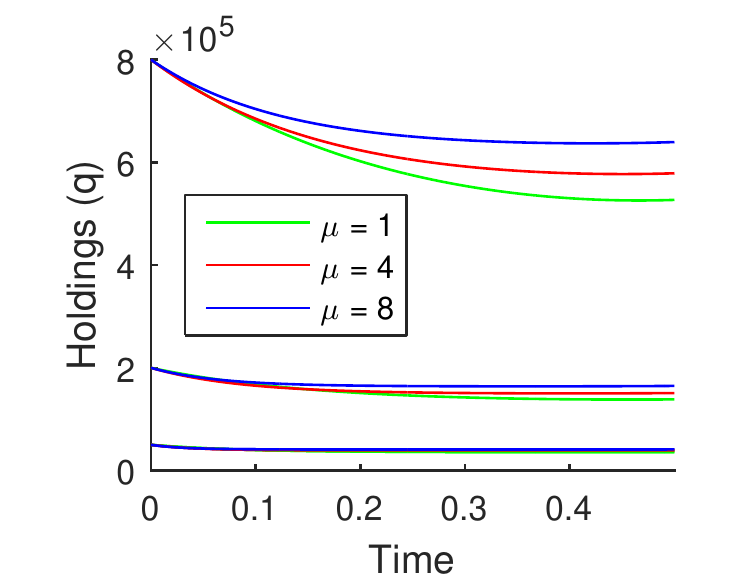}
\caption{} 
\end{subfigure}
\begin{subfigure}[h]{0.45\textwidth}\includegraphics[width=1\columnwidth]{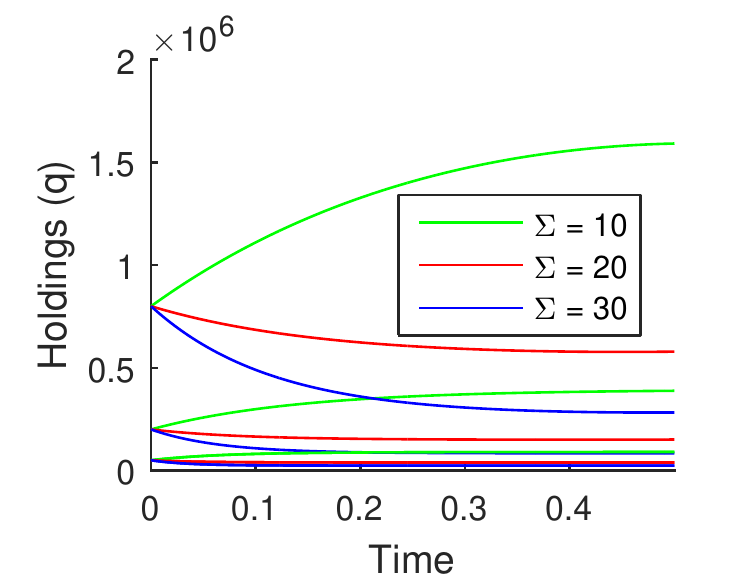}
\caption{} 
\end{subfigure}
\caption{Effects on the benchmark optimal trading curve (red curve) for three firms as one parameter changes upwards (blue curve) and downwards (green curves). (a) shows the effect of changing $\mu$, the mean rate of return of the risky asset; (b) shows the effect of changing $\Sigma$, the volatility of the risky asset.  The vertical axis shows $q$, the amount of the risky asset being held at any time during the trading period. Both figures were computed using a trading period of half a year, while maintaining a probability of default of $1\%$ for all trading curves. }
\label{mu_Sigma_plot}

\end{figure}

\begin{figure}[h]
\centering
\begin{subfigure}[h]{0.45\textwidth}
\includegraphics[width=1\columnwidth]{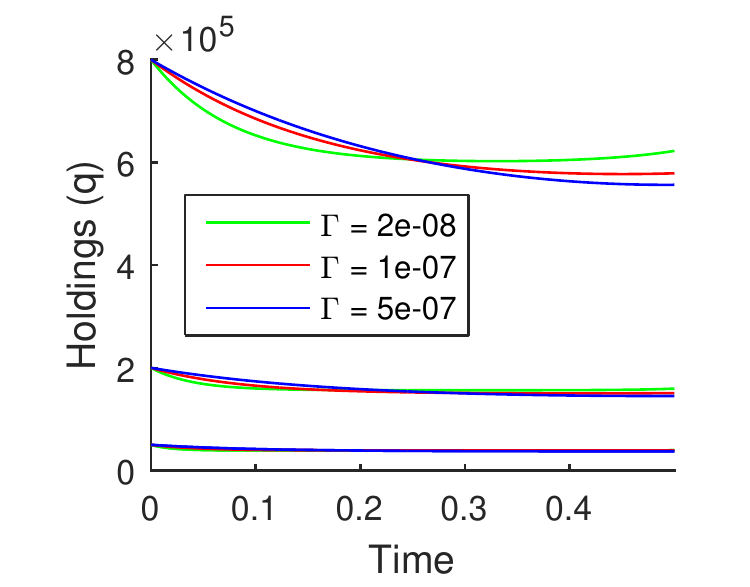}
\caption{}
\end{subfigure}
\begin{subfigure}[h]{0.45\textwidth}
\includegraphics[width=1\columnwidth]{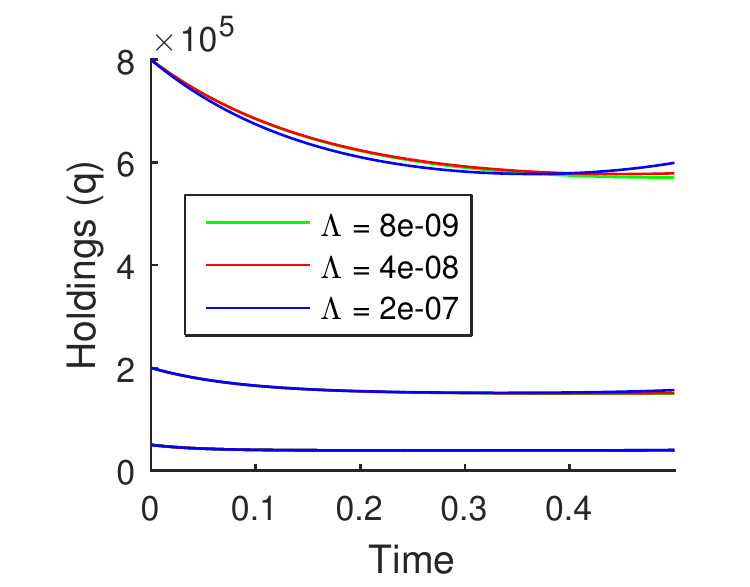}
\caption{}
\end{subfigure}
\caption{Effects on the benchmark optimal trading curve (red curve) for three firms as one parameter changes upwards (blue curve) and downwards (green curves). (a) shows the effect of changing the  temporary impact parameter $\Gamma$; (b)  shows the effect of changing the  permanent  impact parameter $\Lambda$.   The vertical axis shows $q$, the amount of the risky asset being held at any time during the trading period. Both figures were computed using a trading period of half a year, while maintaining a probability of default of $1\%$ for all trading curves. }
\label{Gamma_Lambda_plot}
\end{figure}

 To better understand the properties of the optimal investment strategies that result from our method, we now investigate how the three hypothetical firms' optimal trading in the single asset case compare as important model parameters are varied away from the benchmark parameters of Table \ref{tab:Parameters}. Figures \ref{mu_Sigma_plot} and \ref{Gamma_Lambda_plot} summarize the results of four experiments, and show how the firms' optimal trading strategies over the time period $[0,1/2]$ years change as the asset rate of return, asset volatility, temporary market impact and permanent market impact are made to vary one at a time. In each figure, the red curve denotes the benchmark parametrization, while the other two curves show the result as one specific parameter is varied upwards (blue curve) and downwards (green curves). 
 
 One point needs to be reiterated: for each choice of a set of parameters excluding the risk aversion parameter $\lambda$, $\lambda$ is computed to ensure that the firm's default probability (DP) is exactly 1\%. Thus each curve in these figures corresponds to a different value of $\lambda$. 
 
The effect  on the optimal strategy of varying the asset rate of return $\mu$ and volatility $\Sigma$ is shown in Figure \ref{mu_Sigma_plot}. It is not a surprise to observe   that the optimal strategy will include more of the risky asset as the rate of return is raised, or as the volatility is lowered. There is a threshold value of $\Sigma$ below which the firm switches from sell strategies to  buy strategies. Although not shown in the graph, one finds the reverse is the case for $\mu$.  Finally, the velocity of selling strategies seems to retain a similar shape over time under these variations. Each of these observations are borne out by more extensive investigations of the dependence on these parameters. 

In Figure \ref{Gamma_Lambda_plot}a,   the main effect of decreasing temporary impact $\Gamma$ is seen to be to move more quickly to the final holding level early in the period. This can be understood as a change in the optimal balance between reducing temporary impact costs and price uncertainty due to the asset volatility.  To a lesser extent, one also sees in these examples that the level of the final holdings decreases slightly as $\Gamma$ increases.

The effect of permanent impact $\Lambda$ on the strategy is more subtle.  From Figure \ref{Gamma_Lambda_plot}b, a higher permanent impact  parameter $\Lambda$ leads to an optimal strategy ending with a higher holding level. It also causes more curvature for the trading strategies, especially towards the closing time where all trading curves seem to have positive slope. Indeed, directly from  \eqref{optimalq}, the general formula  for the trading velocity, one verifies that at the close of the period $\frac{dq}{du}\mid_{u=T}=\frac12{\Gamma^{-1}\Lambda}$. This means that as long as $\Lambda$ is positive, every trader holding long positions, whether leveraging up or down, will always end the period by buying more shares. The reason is because  permanent impact gives any trader a small opportunity to push the asset prices in a favourable direction at the last moment. We call this the {\it Ponzi property} of our market impact model: the gains it implies cannot be converted to cash without bursting the small price bubble the trader has created. 

\subsection{Small Market Impact}
\label{smallimpact}
\begin{figure}[h]
\centering
\begin{subfigure}[h]{0.45\textwidth}

\includegraphics[width=1\columnwidth]{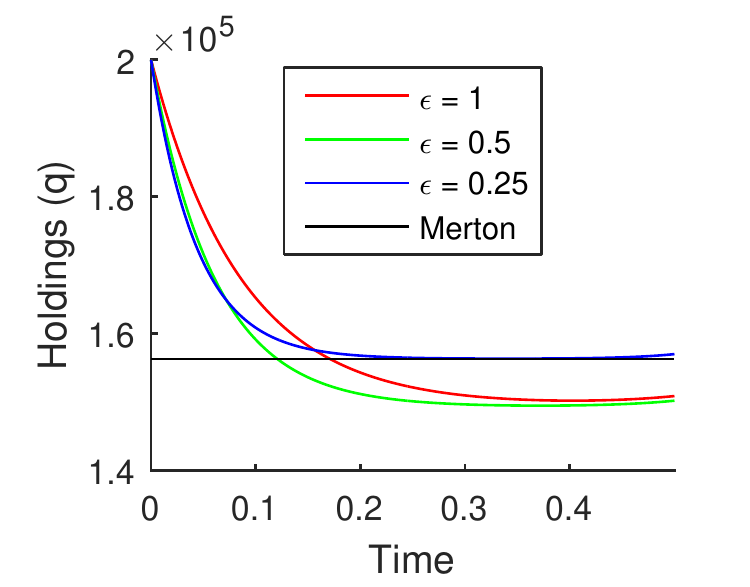}
\caption{$\Lambda=\epsilon\Lambda_1, \Gamma=\epsilon\Gamma_1$}
\end{subfigure}
\begin{subfigure}[h]{0.45\textwidth}
\includegraphics[width=1\columnwidth]{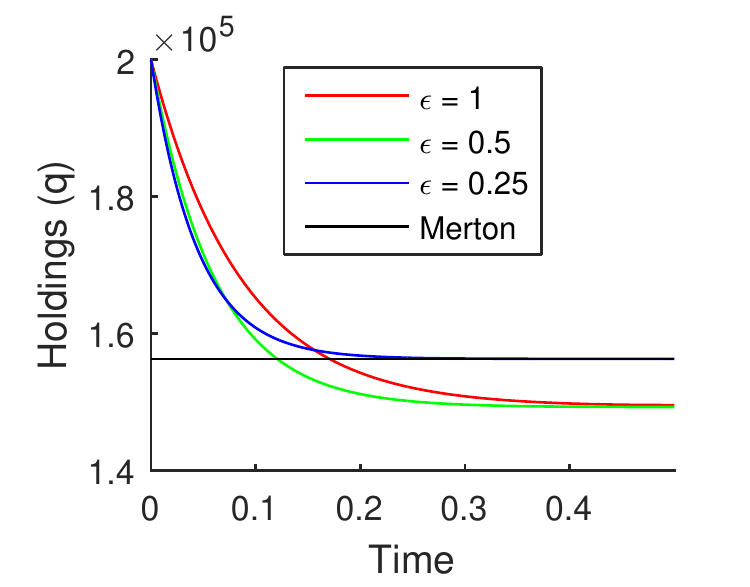}
\caption{$\Lambda=0, \Gamma=\epsilon\Gamma_1$}

\end{subfigure}
\caption{The behaviour of the optimal trading strategy for a decreasing sequence of market impact parameters as described in Section \ref{smallimpact}. They show convergence to the constant Merton solution.}
\label{Merton_plot}
\end{figure}

The perturbative analysis of Section \ref{perturb} provides an alternative framework for understanding the effect of permanent and temporary market impact. We investigate the middle-sized firm with $q_0=2\times 10^5$ and market impact parameters $\Gamma(\epsilon)=\epsilon \times 10^{-7}, \ \Lambda(\epsilon)=\epsilon \times 4 \times 10^{-8}$ for a sequence of values $\epsilon_n=10^{-n}, n=0,1,\dots $ approaching zero. Figure \ref{Merton_plot}(a) shows how the optimal strategies converge for $\epsilon\to 0$ to the constant Merton solution for $u\in(t, T)$, but show rapid transient effects for $u$ near both endpoints. The small Ponzi effect near $u=T$ can be turned off by taking $\Lambda(\epsilon)=0$, as shown in Figure \ref{Merton_plot}(b).

These figures suggest that for reasonable parameter values and small market impact, our model will deliver strategies that are effectively similar to the Merton solution. The observed relationship  between the optimal strategies and the Merton solution, valid  for small market impact,  actually remains true for intermediate levels of market impact such as our benchmark parametrizations. One observes in Figures \ref{mu_Sigma_plot} and \ref{Gamma_Lambda_plot} that all strategies tend to flatten as $u$ approaches $T$, albeit with a small Ponzi effect at the end of the period. It will be well worth studying the extent that the value of the holdings at which the strategy flattens is well approximated by the Merton solution. As the market impact parameters decrease, the flat portion of the curve becomes wider, and closer to the Merton solution.

\begin{figure}[h]
\centering
\begin{subfigure}[h]{0.45\textwidth}
\includegraphics[width=1\columnwidth]{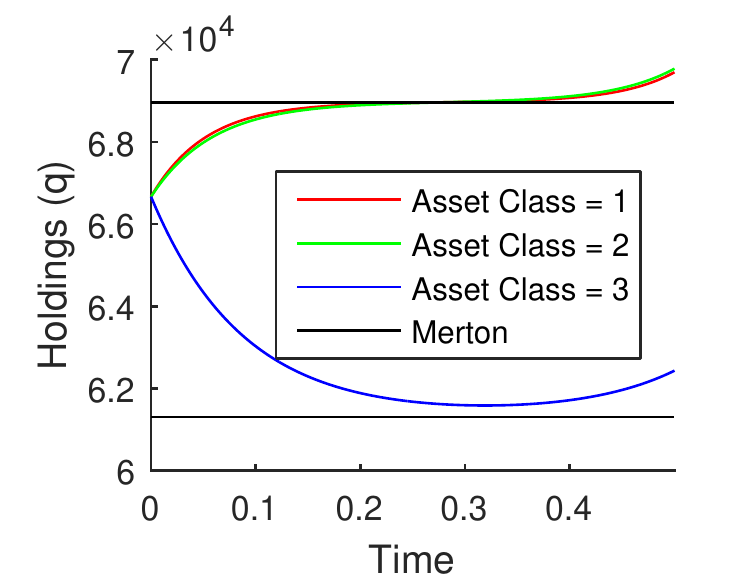}
\caption{3 Uncorrelated Assets}
\end{subfigure}
\begin{subfigure}[h]{0.45\textwidth}
\includegraphics[width=1\columnwidth]{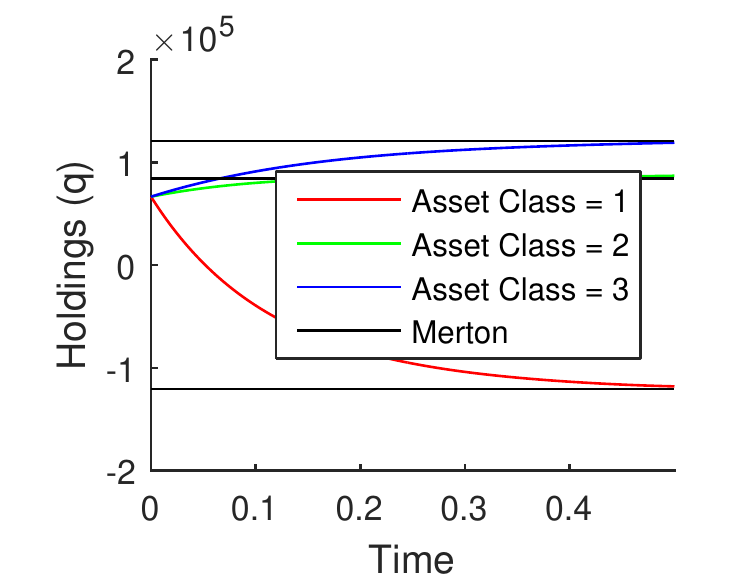}
\caption{3 Positively Correlated Assets}
\end{subfigure}
\caption{The behaviour of the optimal three asset trading strategy in the uncorrelated and correlated cases, when compared to the Merton solution.}
\label{MultiAsset_plot}
\end{figure}

An  analysis similar to that of Section \ref{perturb} allows us to understand the multi--asset investment problem in the small market impact regime. Figure \ref{MultiAsset_plot}, we used the standard asset parameters as Asset 2, with Asset 1 being the asset with the lower perturbed parameters from the previous cases, and Asset 3 having the higher perturbed parameters from the previous cases. Figure \ref{MultiAsset_plot}(a), compares the uncorrelated case to the Merton solution. Figure \ref{MultiAsset_plot}(b), compares the case of constant pairwise correlation $\rho=0.5$ to the Merton solution. In both cases we can see the behaviour similar to the one asset case above. It should be noted that unlike the single asset case, a hedging strategy can be utilized for when multiple assets are available, hence short selling of a illiquid asset class can be optimal.

We observe again in the multi-asset problem that when the market impact is small, the general optimal strategy is close to the Merton solution.

\subsection{Bounded Optimal Trading Strategies}
We have seen in Section \ref{StratProps} that in the single asset case, positive $\Lambda$ creates the Ponzi property that gives any trader an opportunity to push the price in their favour near the end of the period. Case 1 of Proposition \ref{singleassetprop} shows that as long as $\Lambda<\sqrt{2\lambda\Gamma}\Sigma$, the optimal strategies computed over any finite period $[0,T]$ remain bounded. However, when
$\Lambda>\sqrt{2\lambda\Gamma}\Sigma$, Case 3 of Proposition \ref{singleassetprop}  implies that for the period $[0,T^*]$ with $T^*=\tilde K/D$,  the optimal strategy $q^*(u)$ and the value function $W$ both blow up at $u=0$. 

Similar possibilities arise in the multi-asset investment problem.  As $\Lambda$ increases, eventually the matrix function $U(t)$ becomes singular for some finite $t=T^*$. Again, one then finds  that for the period $[0,T^*]$,  the optimal strategy $q^*(u)$ and the value function $W$ both blow up at $u=0$.

\section{Remarks and Conclusions}

The three hypothetical financial institutions studied in Section \ref{NumInv} face a typical investment problem, namely to maximize their return on equity subject to an upper bound on the downside risk, which is defined here as the probability of default. We have presented an analytically tractable version of the optimal portfolio problem that can be justified three different ways: as utility optimization, as mean-variance optimization and as mean-default probability optimization. Numerical evidence shows that the solutions generated by the method have desirable and interesting features. Perhaps most importantly, we have learned that these strategies closely track the classic Merton solution arising in the zero market impact model. 

The three benchmark firms have efficient frontiers shown in Figure \ref{EF_plot} that quantify by how much their rate of return will increase if they raise their tolerance to default. We have observed that optimal trading strategies that account for market impact tend to move over the trading period toward the Merton solution. If they are initially close to the Merton solution, they will tend to remain close, which means the Merton solution is robust to perturbations. The speed of approach increases as the temporary impact parameter $\Gamma$ decreases. In addition, the main effect of the permanent impact $\Lambda$ is the Ponzi property that is manifested by some amount of buying near the end of the period. This Ponzi effect is typically small, but as Proposition \ref{singleassetprop} shows, it will dominate the character of the solution when $\Lambda$ becomes large enough to cause an asset price bubble. 

Left to themselves, there is little incentive for such FIs to limit risk seeking.  By choosing a low value of $\lambda$, or equivalently, accepting a high leverage ratio, they can achieve a high rate of return on capital. Since lower temporary impact and higher permanent impact are both relatively more advantageous to larger firms, one has situations where large firms implement aggressive Ponzi style strategies. In scenarios where the assets perform badly, there is a likelihood of serious asset price feedback that may adversely affect other financial institutions holding common assets. Such asset price feedback, both bubbles and bursts, has been identified in the literature, notably \citet{cifuentes2005liquidity}, as  a critical channel of systemic risk, popularly known as the {\it asset fire sale channel}. One application of our model, yet to be explored in detail, will be its use to specify the natural behaviour of the banks and financial institutions in a large financial system, and then to see how systemic risk measures are affected by asset fire sales due to market impact. In this systemic risk context, it will also be important to introduce the effects of funding illiquidity by modelling the stochastic nature of deposits.

If large banks were permitted to act in their own self interest without regard to their systemic effects, they would pose an unacceptable threat to financial stability. For that reason, all banks are subjected to a regime of strict financial regulation, of which the most important are limits to their capital asset ratio and liquidity coverage ratio. Under such regulatory constraints, FIs' investment strategies will differ dramatically from the optimal strategies produced in the present paper. The optimal behaviour of such regulated financial institutions will be the target of future modelling studies.

\bibliographystyle{abbrvnat}
\bibliography{PortfoliosIlliquidAssets_HurdShaoTuan}

\begin{thebibliography}{19}
\providecommand{\natexlab}[1]{#1}
\providecommand{\url}[1]{\texttt{#1}}
\expandafter\ifx\csname urlstyle\endcsname\relax
  \providecommand{\doi}[1]{doi: #1}\else
  \providecommand{\doi}{doi: \begingroup \urlstyle{rm}\Url}\fi

\bibitem[Almgren and Chriss(2001)]{AC}
R.~Almgren and N.~Chriss.
\newblock Optimal execution of portfolio transactions.
\newblock \emph{Journal of Risk}, 3:\penalty0 5--40, 2001.

\bibitem[Almgren and Lorenz(2007)]{almgren2007adaptive}
R.~Almgren and J.~Lorenz.
\newblock Adaptive arrival price.
\newblock \emph{Trading}, 2007\penalty0 (1):\penalty0 59--66, 2007.

\bibitem[Almgren(2003)]{almgren2003optimal}
R.~F. Almgren.
\newblock Optimal execution with nonlinear impact functions and
  trading-enhanced risk.
\newblock \emph{Applied Mathematical Finance}, 10\penalty0 (1):\penalty0 1--18,
  2003.

\bibitem[Brown et~al.(2010)Brown, Carlin, and Lobo]{brown2010optimal}
D.~B. Brown, B.~I. Carlin, and M.~S. Lobo.
\newblock Optimal portfolio liquidation with distress risk.
\newblock \emph{Management Science}, 56\penalty0 (11):\penalty0 1997--2014,
  2010.

\bibitem[Brunnermeier and Pedersen(2009)]{BrunPede09}
M.~K. Brunnermeier and L.~H. Pedersen.
\newblock Market liquidity and funding liquidity.
\newblock \emph{Review of Financial Studies}, 22\penalty0 (6):\penalty0
  2201--2238, 2009.

\bibitem[Cartea and Jaimungal(2015)]{cartea2015optimal}
{\'A}.~Cartea and S.~Jaimungal.
\newblock Optimal execution with limit and market orders.
\newblock \emph{Quantitative Finance}, 15\penalty0 (8):\penalty0 1279--1291,
  2015.

\bibitem[Cifuentes et~al.(2005)Cifuentes, Ferrucci, and
  Shin]{cifuentes2005liquidity}
R.~Cifuentes, G.~Ferrucci, and H.~S. Shin.
\newblock Liquidity risk and contagion.
\newblock \emph{Journal of the European Economic Association}, 3\penalty0
  (2-3):\penalty0 556--566, 2005.

\bibitem[Davis and Norman(1990)]{davis1990}
M.~H. Davis and A.~R. Norman.
\newblock Portfolio selection with transaction costs.
\newblock \emph{Mathematics of Operations Research}, 15\penalty0 (4):\penalty0
  676--713, 1990.

\bibitem[Gatheral and Schied(2011)]{Gatheral}
J.~Gatheral and A.~Schied.
\newblock Optimal trade execution under geometric brownian motion in the
  almgren and chriss framework.
\newblock \emph{International Journal of Theoretical and Applied Finance},
  14\penalty0 (03):\penalty0 353--368, 2011.

\bibitem[Hurd et~al.(2016)Hurd, Shao, and Tran]{HurdShaoTuan16a}
T.~R. Hurd, Q.~H. Shao, and T.~Q. Tran.
\newblock Review of portfolio strategies in illiquid markets.
\newblock available at, June 2016.

\bibitem[Markowitz(1952)]{Mark52}
H.~Markowitz.
\newblock Portfolio selection.
\newblock \emph{The Journal of Finance}, 7\penalty0 (1):\penalty0 77--91, 1952.

\bibitem[Merton(1969)]{Merton69}
R.~C. Merton.
\newblock Lifetime portfolio selection under uncertainty: the continuous--time
  model.
\newblock \emph{Rev. Econom. Statist.}, 51:\penalty0 247--257, 1969.

\bibitem[Moazeni et~al.(2013)Moazeni, Coleman, and Li]{moazeni2013optimal}
S.~Moazeni, T.~F. Coleman, and Y.~Li.
\newblock Optimal execution under jump models for uncertain price impact.
\newblock \emph{Journal of Computational Finance}, 16\penalty0 (4):\penalty0
  1--44, 2013.

\bibitem[Perold(1988)]{perold1988implementation}
A.~F. Perold.
\newblock The implementation shortfall: Paper versus reality.
\newblock \emph{The Journal of Portfolio Management}, 14\penalty0 (3):\penalty0
  4--9, 1988.

\bibitem[Pham and Tankov(2008)]{Pham}
H.~Pham and P.~Tankov.
\newblock A model of optimal consumption under liquidity risk with random
  trading times.
\newblock \emph{Mathematical Finance}, 18\penalty0 (4):\penalty0 613--627,
  2008.

\bibitem[Schied and Sch{\"o}neborn(2009)]{schied2009risk}
A.~Schied and T.~Sch{\"o}neborn.
\newblock Risk aversion and the dynamics of optimal liquidation strategies in
  illiquid markets.
\newblock \emph{Finance and Stochastics}, 13\penalty0 (2):\penalty0 181--204,
  2009.

\bibitem[Schied et~al.(2010)Schied, Sch{\"o}neborn, and
  Tehranchi]{schied2010optimal}
A.~Schied, T.~Sch{\"o}neborn, and M.~Tehranchi.
\newblock Optimal basket liquidation for cara investors is deterministic.
\newblock \emph{Applied Mathematical Finance}, 17\penalty0 (6):\penalty0
  471--489, 2010.

\bibitem[Sch{\"o}neborn(2008)]{Schoneborn}
T.~Sch{\"o}neborn.
\newblock Trade execution in illiquid markets: Optimal stochastic control and
  multi-agent equilibria. phd thesis.
\newblock 2008.

\bibitem[Zhang(2014)]{Zhang}
T.~Zhang.
\newblock \emph{Nash Equilibria in Market Impact Models: Differential Game,
  Transient Price Impact and Transaction Costs, PhD Thesis}.
\newblock PhD thesis, Mannheim, Universit{\"a}t Mannheim, Diss., 2014, 2014.

\end{thebibliography}

\appendix
\section{Appendix: Proofs of Main Results}\label{A1}

{\noindent} {\bf Proof of Theorem \ref{Optimal}:\ } In this proof we fix $T$ to be finite. The existence of a maximal $T^*$ is a consequence of solving \eqref{MV_optimal}, which is analyzed in the proof of Proposition \ref{RicattiEqn}. The Hamilton-Jacobi-Bellman (HJB) equation associated to \eqref{DPP1} arises from the DPP by assuming Markov controls $v_t=v(t,T,X_t,q_t)$ and value function $W_t:=W(t,T,X^v_t,q^v_t)$ for deterministic functions $v,W$. For simplicity of exposition, we have omitted the potential for dependence on the stock price $S_t$: the standard verification result used at the end of this argument shows this is consistent. 

Under these assumptions, the DPP implies that $-e^{-\lambda  W(t,T,X^v_t,q^v_t)}$ is a supermartingale for all $v$ and a martingale for the optimal $v^*$, which leads to the HJB equation for $W$ 
\begin{eqnarray*}
&&\partial_t W+ \partial_XW q'\mu +\frac{1}{2} q'\Sigma\Sigma' q [\partial_{XX}^2 W-\lambda(\partial_XW)^2]\\
&&\hspace{1.5in}+ \ {\sup}_v [(\partial_qW'+\partial_XWq'\Lambda)v-v'\Gamma v \partial_XW]=0.\\
&&W(T,T, X,q)= X.
\end{eqnarray*}
The ansatz $W(t, T,  X, q)=X+V(t,T, q)$ leads to the equation for $V$
\begin{eqnarray*}
&&\partial_t V+ q'\mu -\frac{\lambda}{2} q'\Sigma\Sigma' q + \ {\sup}_v [(\partial_qV'+q'\Lambda)v-v'\Gamma v]=0.\\
&&V(T, T,q)=0.
\end{eqnarray*}
The optimal feedback control is thus $v^*=\frac{\Gamma^{-1}}{2}(\partial_qV'+\Lambda q),$ which is independent of $X$ and  the price process, and hence deterministic. Using this control leads to
\be\label{HJB_V} \partial_t V+q'\mu-\frac{\lambda}{2} q'\Sigma\Sigma' q+\frac 14 (\partial_qV+\Lambda q)'\Gamma^{-1}(\partial_qV+\Lambda q)=0\ .\ee
 As we will shortly see in the proof of Proposition \ref{RicattiEqn}, this ODE has a unique smooth solution which is deterministic, over any finite time interval $[t,T]$ for $T$ less than a possibly infinite maximal $T^*$. Therefore, by the classical verification theorem, we have $W=\tilde{W}$ and the other statements of the theorem follow.\\
\noindent\qedsymbol

\bigskip

{\noindent} {\bf Proof of Proposition \ref{RicattiEqn}:\ } By Theorem \ref{Optimal} , the value function for Merton's problem over $[t,T]$ has the form $W(t, T, X,q)=X+V(t,T, q),$ where $V$ satisfies the ODE \eqref{HJB_V}. This ODE and the form \eqref{Vform} leads to  Riccati equations with initial conditions for $A, B, C$
\begin{eqnarray}
\label{riccatiA2}\partial_\tau A&&-\frac 14(A+A'+ \Lambda)\Gamma^{-1}(A+A'+\Lambda)+\frac{\lambda}{2}\Sigma\Sigma'=0,\quad A(0)=0\\
\label{riccatiB2}\partial_\tau B&&-\frac 12B\Gamma^{-1}(A+A'+\Lambda)-\mu'=0,\quad B(0)=0\\
\label{riccatiC2}\partial_\tau C&&-\frac 14B\Gamma^{-1}B'=0, \quad C(0)=0.
\end{eqnarray}
Notice that if $A$ is a solution of (\ref{riccatiA}), then so is $A'$: By the uniqueness theorem for solutions of ODEs, $A=A'$ and therefore $A$ is symmetric.

\noindent\qedsymbol

\bigskip

{\noindent} {\bf Proof of Theorem \ref{closedform}:\ } 
Part 1: Note that  $\frac{\lambda}{2}\Gamma^{-1/2}\Sigma\Sigma'\Gamma^{-1/2}$ is  positive definite and define $D$ to be its symmetric square root. If
\[\tilde A(\tau):=\Gamma^{-1/2}(A(\tau)+\Lambda/2)\Gamma^{-1/2}\]
then \eqref{riccatiA} becomes
\begin{equation}
\label{riccatiA3}
\partial_\tau\tilde A-\tilde A^2+D^2=0,\quad \tilde A(0)=E:=\Gamma^{-1/2}(\Lambda/2)\Gamma^{-1/2}\ .
\end{equation}
One can now check that the solution to \eqref{riccatiA3} has the form $\tilde A=VU^{-1},$ where $U, V$ satisfy the following  linear ODE with terminal condition
\[
\left[ \begin{array}{c} \partial_\tau U \\ \partial_\tau V \end{array} \right] = \left[ \begin{array}{cc} 0 & - \mathbbm{1} \\ -D^2 & 0 \end{array} \right] \times \left[ \begin{array}{c} U \\ V \end{array} \right],\quad \left[ \begin{array}{c} U(0) \\ V(0) \end{array} \right]=\left[ \begin{array}{c} \mathbbm{1} \\ E \end{array} \right] \ .
\]
By block-diagonalization using
\[
 Q= \left[ \begin{array}{cc} \mathbbm{1} & \mathbbm{1} \\ D & -D \end{array} \right] ,\quad  Q^{-1}= \frac 12\left[ \begin{array}{cc} \mathbbm{1} & D^{-1} \\ \mathbbm{1} & -D^{-1} \end{array} \right]  
\] one finds
\[
 \left[ \begin{array}{cc} 0 &- \mathbbm{1} \\ -D^2 & 0 \end{array} \right]= Q\left[ \begin{array}{cc} -D & 0 \\ 0 & D \end{array} \right]Q^{-1}  
\]
and therefore, the solution of the matrix ODE is
\[
\left[ \begin{array}{c} U(\tau) \\ V(\tau) \end{array} \right] = Q\left[ \begin{array}{cc} e^{-D\tau} & 0 \\ 0 & e^{D\tau} \end{array} \right]Q^{-1} \times \left[ \begin{array}{c} \mathbbm{1} \\ E \end{array} \right] \ .
\]
From the explicit forms 
\begin{eqnarray}
U(\tau)&&=\ {\cosh}(D\tau)-\ {\sinh}(D\tau)D^{-1}E\\
V(\tau)&&=-\ {\sinh}(D\tau)D+\ {\cosh}(D\tau)E
\end{eqnarray}
one finds $A(\tau)=\Gamma^{1/2}(\tilde{A}(\tau)-E)\Gamma^{1/2}$ where
\be 
\tilde A(\tau)=[-\ {\sinh}(D\tau)D+\ {\cosh}(D\tau)E][\ {\cosh}(D\tau)-\ {\sinh}(D\tau)D^{-1}E]^{-1}.
\ee

The Riccati equation \eqref{riccatiB} for B can be solved by noting that $\tilde B= B\Gamma^{-1/2}$ solves the ODE 
$$\partial_\tau\tilde B-\tilde B\tilde A-\mu'\Gamma^{-1/2}=0.$$ 
Since $\partial_\tau U=-\tilde AU$, we find
$ \partial_\tau(\tilde B U)=(\partial_\tau\tilde B-\tilde B \tilde A)U=\mu'\Gamma^{-1/2}U$
which can be integrated to give
  $\tilde B(\tau)U(\tau)=\mu'\Gamma^{-1/2}(\int^\tau_0 U(s)ds)$ and thus
  \[B(\tau)=\mu'\Gamma^{-1/2}\left(\int_0^\tau U(s)ds\right)U^{-1}(\tau)\Gamma^{1/2}\ .
  \]
 It is straightforward that
$\int^\tau_0 U(s)ds=D^{-2}[E-V(\tau)] $
which gives the desired formula
$$B(\tau)=\mu'\Gamma^{-1/2}D^{-2}[E-V(\tau)]U^{-1}(\tau)\Gamma^{1/2}\ .$$

In a similar fashion, one finds
\begin{eqnarray}
C(\tau)&&=\frac 14\int_0^\tau B(s)\Gamma^{-1} B'(s)ds\\
&&=\frac 14\overline{\mu}'\left(\int_0^\tau(E-V(s))(U'(s)U(s))^{-1}(E-V'(s))ds\right)\overline{\mu},
\end{eqnarray}
where $\overline{\mu}:=D^{-2}\Gamma^{-1/2}\mu.$

Part 2: This part is straightforward.

Part 3: From part 4 of  Theorem \ref{Optimal}, the optimal control $q^*(u)$ over the period $[t,T]$ solves
$$\partial_u q -\Gamma^{-1}(A(T-u)+\Lambda/2)q=\frac 12\Gamma^{-1}B'(T-u)$$
When this linear ODE is multiplied on the left by the integrating factor $U^{-1}(T-u) \Gamma^{1/2}$, the left-hand side becomes an exact derivative:
$$\partial_u \left[U^{-1}(T-u)\Gamma^{1/2} q\right]=U^{-1}(T-u)\Gamma^{1/2}\times \frac 12\Gamma^{-1}B'(T-u)\ .$$
 Integration of this equation over $[t,u]$ gives
 $$ U^{-1}(T-u)\Gamma^{1/2} q(u)-U^{-1}(T-t)\Gamma^{1/2} q= \frac 12 \int^u_t U^{-1}(T-r)\Gamma^{-1/2} B'(T-r)dr$$ which leads to the desired formula. 
 
Part 4:  The Variance is calculated directly as follows
$$\text{Var}_t(X_T^*)=\int_t^Tq^*(s)'\Sigma\Sigma'q(s)^*ds=q'L(T-t)q+M(T-t)q+N(T-t)\ .$$

Rewrite $q^*(u)={\tilde U}{(T-u)}{\tilde U}^{-1}{(T-t)}q+\frac 12{\tilde U}{(T-u)}I(u)$, where $\tilde U(T-u):=\Gamma^{-1/2}U(T-u)$ and  $I(u):=\int_t^u{\tilde U}^{-1}(T-r) \Gamma^{-1}B(T-r)dr$. Explicit forms for $L, M, N$ are calculated as follows.
\begin{eqnarray*}
L(T-t)=(\tilde U^{-1}(T-t)')\Bigl(\int_t^T\tilde U(T-r)'\Sigma\Sigma'\tilde U(T-r)dr\Bigr)\tilde U^{-1}(T-u)\ .
\end{eqnarray*}
 By using Fubini's formula, we have
 \begin{eqnarray*}
 M'(T-t)&&=\int_t^T{\tilde U^{-1}}{(T-t)'}{\tilde U}{(T-r)'}\Sigma\Sigma'{\tilde U}{(T-r)}I(r)dr\\
 &&=\int_t^T\Bigl(\int_s^T{\tilde U^{-1}}{(T-t)'}{\tilde U}{(T-r)'}\Sigma\Sigma'{\tilde U}{(T-r)}dr\Bigr){\tilde U}{(T-s)}^{-1}\Gamma^{-1}B(T-s)ds\\
 &&={\tilde U}{(T-t)'}^{-1}\int_t^T{\tilde U}{(T-s)}L(T-s)\Gamma^{-1}B(T-s)ds\ .
 \end{eqnarray*}
 Similarly
\begin{eqnarray*}
N(T-t)&&=\frac{1}{4}\int_t^TI(r)'{\tilde U}{(T-r)}'\Sigma\Sigma'{\tilde U}{(T-r)}I(r)dr\\
&&=\frac{1}{4}\int_t^T\Bigl(\int_s^TI(r)'{\tilde U}{(T-r)}'\Sigma\Sigma'{\tilde U}{(T-r)}dr\Bigr){\tilde U}^{-1}(T-s)\Gamma^{-1}B(T-s)ds\\
&&= \frac{1}{4}\int_t^TM(T-s)\Gamma^{-1}B(T-s)ds\ .
 \end{eqnarray*} 

\noindent\qedsymbol

 \end{document}